\documentclass[twocolumn,nopacs,preprintnumbers,pra,aps,10pt,superscriptaddress]{revtex4}

\usepackage{graphicx,color}% Include figure files
\usepackage{dcolumn}% Align table columns on decimal point
\usepackage{bbm, bm, amsmath, amssymb, hyperref}% bold math
\usepackage[caption=false]{subfig}
\usepackage{pdfsync}

\newcommand{\ket}[1]{|#1\rangle}

\def\>{\rangle}
\def\<{\langle}
\def\E{ {\cal E} }

\def\R{ {\cal R} }

\def\I{ \mathbbm{1} }

\def\d{\mathrm{d}}

\def\lhat{\hat{\lambda}}

\def\psihat{\hat{\psi}}

\def\supp{\mbox{Supp}\,}

\begin{document}

\title{Distinct Quantum States Can Be Compatible with a Single State of Reality\footnote{This article was published by the American Physical Society under the terms of the Creative Commons Attribution 3.0 License. Further distribution of this work must maintain attribution to the authors and the published article's title, journal citation (PRL {\bf 109}, 150404 (2012)), and DOI (\href{http://dx.doi.org/10.1103/PhysRevLett.109.150404}{10.1103/PhysRevLett.109.150404}).}}

\author{Peter G. Lewis}\email{1@physics.org}
\affiliation{Controlled Quantum Dynamics Theory Group, Imperial College London, London SW7 2AZ, United Kingdom}

\author{David Jennings}
\affiliation{Controlled Quantum Dynamics Theory Group, Imperial College London, London SW7 2AZ, United Kingdom}

\author{Jonathan Barrett}
\affiliation{Department of Mathematics, Royal Holloway, University of London, Egham Hill, Egham TW20 0EX, United Kingdom}

\author{Terry Rudolph}
\affiliation{Controlled Quantum Dynamics Theory Group, Imperial College London, London SW7 2AZ, United Kingdom}

\begin{abstract}

Perhaps the quantum state represents information about reality, and not reality directly. Wave function collapse is then possibly no more mysterious than a Bayesian update of a probability distribution given new data. We consider models for quantum systems with measurement outcomes determined by an underlying physical state of the system but where several quantum states are consistent with a single underlying state---i.e., probability distributions for distinct quantum states overlap. Significantly, we demonstrate by example that additional assumptions are always necessary to rule out such a model.

\end{abstract}

\date{\today}

\maketitle

Broadly speaking, physicists today view the quantum state in one of two ways: in correspondence with the real physical state of affairs or as representing only an agent's knowledge or information about some aspect of the physical situation. The latter `epistemic' viewpoint is primarily motivated by the obvious parallel between the quantum process of instantaneous wavefunction collapse, and the classical procedure of instantaneous updating of a probability distribution, both of which occur upon the acquisition of information regarding the outcome of a measurement process. The epistemic view has a long history of illustrious advocates \cite{EPR35,Pop67,Bal70,Pei79,Jay80,Zei99,Cav02,SpekToy}. In recent times a research programme has arisen aimed at not only philosophically justifying the epistemic view, but potentially deriving quantum theory from more primitive considerations about information and/or Bayesian reasoning \cite{Bru01, Fuc10, SpekToy,Lei06,Lei11,Lei11b,Bar09,Har01,Bar07}.

At least two versions of the epistemic view can be distinguished. One is operational: the quantum state represents information about which outcome will occur if a measurement is performed on the system. Measurement itself is treated as a primitive. The other is that the quantum state represents information about some underlying physical state of the system, where this underlying state need not be described by quantum theory.

In his 1935 letter to Schr\"{o}dinger, in an attempt to explain what he really meant in the EPR paper, Einstein writes \cite[][p.\,26]{EA22-047,How06}:
\begin{quotation}But then for the same [real] state of [the system] there are two (in general arbitrarily many) equally justified $\Psi$, which contradicts the hypothesis of a one-to-one or complete description of the real states.
\end{quotation}
Einstein came to this conclusion, which he termed `incompleteness' of the quantum state, using an argument based on what we now call (following Schr\"odinger \cite{Sch35}) \emph{steering} \cite{Har10}.  Note that the question he raises here is \emph{not} whether there are multiple states of reality associated with a single wavefunction (one possible type of incompleteness), but rather whether there are multiple wavefunctions associated with a single real state. A natural way to understand this is as an expression of the second kind of epistemic view above---that a quantum state represents an agent's information about an underlying reality, but is not part of that reality itself.

Unfortunately, as shown later by Bell \cite{Bel66} Einstein's specific argument for incompleteness was based on a false premise (locality). However, a question that stands on its own and is the topic of this article remains: is it even mathematically possible to find an embedding of quantum theory in some deeper theory where the quantum states are not always uniquely determined given the underlying physical state? Following \cite{Har10}, we refer to such a possibility as a $\psi$-\emph{epistemic} interpretation of the quantum state.

Recently, a no-go theorem was proven \cite{PBR} showing that a $\psi$-epistemic interpretation is impossible. It is important to note that a key assumption of the argument is \emph{preparation independence}---situations where quantum theory assigns independent product states are presumed to be completely describable by independently combining the two purportedly deeper descriptions for each system. Here, we will show via explicit constructions that without this assumption, $\psi$-epistemic models can be constructed with all quantum predictions retained. Hence, we show that not only is the `preparation independence' assumption of that particular no-go theorem necessary, but also any similar no-go theorem will require non-trivial assumptions beyond those required for a well-formed ontological model.

One of the most compelling motivations for exploring the $\psi$-epistemic view is the amazing range of phenomena, normally considered uniquely quantum, that can be derived by imposing only a simple principle restricting an agent's knowledge about a presumed underlying reality \cite{SpekToy,hardytoytheory,Bar11}. It is clear that the primary reason these theories do manage to reproduce so many quantum-like phenomena is that the states of knowledge (probability distributions) overlap on a non-trivial subset of the underlying space of `hidden' states, even when the agent's knowledge is maximal (the equivalent of a quantum pure state).

These toy theories do not, however, reproduce \emph{all} quantum states and measurement statistics. About such theories which do reproduce all of quantum theory much less is known---the main examples and constraints are outlined in \cite{Har07}. In \cite{Har04} Hardy showed that the underlying mathematical space of real states on which such theories are defined---the `ontic state' space---must have cardinality at least that of the integers. Spekkens \cite{Spe05} showed the models must be \emph{preparation contextual} in addition to measurement contextual \cite{Koc67}. The manifold dimension of the ontic state space was shown by Montina \cite{Mon08} to be necessarily exponential assuming that the dynamics of the ontic states is Markovian; he then went on to show the intriguing result that it is possible to reduce the manifold dimension by one \cite{Mon11c,Mon11b}. This latter construction results in probability distributions that intersect in the ontic state space, however they do so only on a set of measure 0. The question of economic representation of finite data in such models \cite{Har01,Gal09,Dak08,Weh08} has also received some attention.

In another pair of recent works \cite{Col11a,Col11b}, Colbeck and Renner have argued that, given additional assumptions of experimenters' free choice in choosing measurement settings and no superluminal signaling of the choice at the ontic level, the ontic states must be in one-to-one correspondence with the quantum states. The models of the present Letter are explicitly constructed for single systems only and, if applied to systems composed of multiple separated subsystems, would involve superluminal influences of measurement choices upon ontic variables. Hence, our results are consistent with those of \cite{Col11a,Col11b}.

{\em Formal statement of the problem.--}Following \cite{Har10}, an \emph{ontological model} for a quantum system defines a measure space $\Lambda$, the elements $\lambda$ of which are termed \emph{ontic states}. These should be thought of as the underlying physical states that a system can be in at a given time. A pure quantum state $|\psi\>$ corresponds to an equivalence class of experimental preparations, and is represented within an ontological model as a probability distribution $\mu_\psi(\lambda)$ over $\Lambda$. The distribution $\mu_\psi$ is called an {\em epistemic state}. It represents the information that an agent has about the ontic state of a system, given that it was prepared in a particular way. Dynamics correspond to trajectories of ontic states through $\Lambda$, and these map to dynamical changes in the probability distributions over $\Lambda$. This setting includes the standard Hilbert space description through the trivial assignment of $\Lambda$ as the complex projective space $CP^{d-1}$ (the boundary of the quantum state space), and $\mu_\psi(\lambda)$ being a delta-function distribution centred at $\lambda=|\psi\>\<\psi|$ \cite{Bel95}. It is for this reason the term `ontic' is used---as opposed to `hidden', for example.

Throughout, we shall consider only projective measurements on a finite $d$-dimensional quantum system, described as a string of rank-1 projectors $\Phi=\{ |\phi_0\>\< \phi_0|, \cdots , |\phi_{d-1}\>\< \phi_{d-1}| \}$, where $\sum_k |\phi_k\>\<\phi_k|=\I$. If a measurement is performed, the probabilities for different outcomes are defined by the ontic state $\lambda$ of the system. Hence each measurement is associated with a set of $d$ \emph{response functions} $\{ \xi_{\phi_k} : \Lambda \rightarrow [0,1] \}$, where $\xi_{\phi_k}(\lambda)$ is the probability of obtaining the outcome $ |\phi_k\>\< \phi_k| $ when the ontic state of the system is $\lambda$. The response functions are positive semi-definite and normalized so that $\sum_k \xi_{\phi_k} (\lambda) = 1, \forall \lambda$. If the $\xi_{\phi_k}(\lambda)$ take values only in $\{0,1\}$ the model is \emph{deterministic}. By the Kochen-Specker theorem \cite{Koc67}, a deterministic ontological model for a quantum system must be measurement contextual \cite{Har07} if the system has dimension $d \geq 3$. This means that the response function $\xi_{\phi_k}$ depends on the complete set of co-measured projectors in $\Phi$. We do not indicate this dependence for notational simplicity.

An ontological model is successful in explaining quantum measurement statistics for measurement $\Phi$ and preparation $\ket{\psi}$ if and only if it is the case that $\forall k$:
\begin{equation}
\begin{aligned}
\int_\Lambda  \xi_{\phi_k}(\lambda) \d\mu_\psi&= \int_\Lambda  \mu_\psi(\lambda)\xi_{\phi_k}(\lambda) \d \lambda =|\<\phi_k|\psi\>|^2.
\end{aligned}
\end{equation}

\begin{figure}
\centering
\setlength{\unitlength}{\columnwidth}
\begin{picture}(1,0.55)
\put(0.05,0.05){\includegraphics[width=0.9\columnwidth]{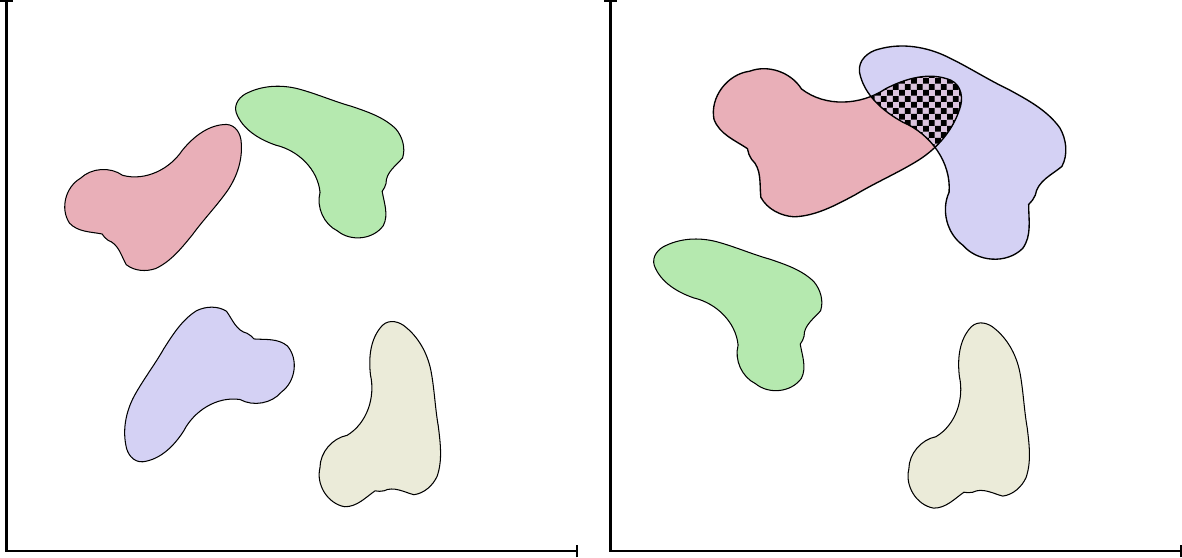}}
\put(0.04,0.49){\(\Lambda\)}
\put(0.5,0.49){\(\Lambda\)}
\put(0.26,0.26){\(\supp(\mu_\psi)\)}
\put(0.250,0.25){\vector(-1, -1){0.06}}
\put(0.07,0.42){\(\supp(\mu_\phi)\)}
\put(0.1,0.4){\vector(1, -1){0.08}}
\put(0.68,0.45){\vector(1, -1){0.06}}
\put(0.66,0.46){\(\mathcal{E}\)}
\end{picture}
\caption{Schematic of two ontic state spaces in different ontological models. The supports of epistemic states associated with four quantum states are shown. In (a) each ontic state is in the support of the epistemic state for at most one \(\psi\): the model is \(\psi\)-ontic. In (b) those ontic states in the highlighted `epistemic region' \(\mathcal{E}\) on the right do not uniquely identify a quantum state, and could result from either of the associated preparation procedures: the model is \(\psi\)-epistemic.\label{epistemic_model}} 
\end{figure}
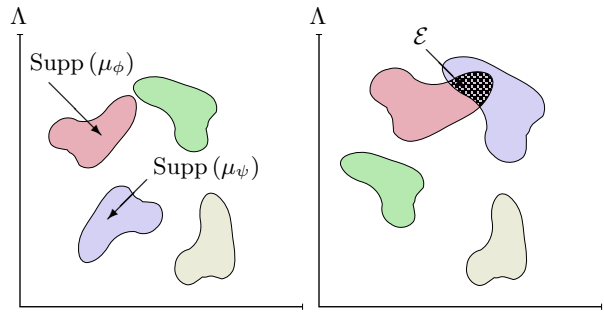

With these basic ingredients in place, an ontological model is \textit{$\psi$-epistemic} if at least two distinct quantum states $|\psi_1\>$ and $|\psi_2 \>$ are described by distributions $\mu_{\psi_1}$ and $\mu_{\psi_2}$ such that the intersection of their supports has non-zero measure, shown schematically in Fig.~\ref{epistemic_model}. An ontological model is \emph{$\psi$-ontic} otherwise \cite{Har10,Spe05}. The idea here is that in a $\psi$-ontic model, the quantum state is uniquely determined by the ontic state $\lambda$, since for any $\lambda$, there is only one $|\psi\rangle$ such that $\lambda$ is contained in the support of $\mu_{\psi}$. In this case, although the quantum state $|\psi\rangle$ plays an epistemic role in defining a distribution $\mu_{\psi}$, the quantum state is also a function of the ontic state, hence can justifiably be thought of as a physical property of the system. Informally, the whole of the quantum state $|\psi\rangle$ is `written into' the real state of affairs.

In a $\psi$-epistemic model, on the other hand, there are at least some circumstances in which two different quantum states describe systems in the same ontic state $\lambda$. In this case, it is defensible to claim that the quantum state `merely' represents an agent's information. A stronger definition of the term $\psi$-epistemic would require that for \emph{any} non-orthogonal states,  $|\psi_1\>$ and $|\psi_2 \>$, the distributions $\mu_{\psi_1}$ and $\mu_{\psi_2}$ overlap. In this Letter, we stick with the weaker definition above.

\emph{The original Bell model.--}As a counterexample to `von-Neumann's silly assumption'  Bell \cite{Bel66} described a simple ontological model capable of describing a quantum system of dimension $d$. The ontic state space consists of pairs $(|\lambda\rangle, x)$, where $|\lambda\rangle$ is an element of $CP^{d-1}$ and $x$ is an element of the interval $[0,1]$. For the moment consider $d=2$. The ontic state space is then isomorphic to $\Lambda = S^2 \times [0,1] $, where $S^2$ is the two-dimensional Bloch sphere. The quantum state $|\psi\rangle$ corresponds to a distribution
$\mu_{\psi} (\lhat , x) = \delta( \lhat - \psihat)$, where the convention here, and in what follows, is to denote the unit Bloch vector associated to $|\psi\>$ as $\psihat$. This choice of distribution is uniform over the subset $\{ (\psihat,x): 0 \le x \le 1 \}$. The model is deterministic, with response functions given by
\begin{eqnarray}\label{BellIndicator}
\xi_{\phi_k} (\lhat, x) &=& \Theta \left [ (|\<\lambda|\phi_0\>|^2 - x) (-1)^k \right ],
\end{eqnarray}
where $\Theta$ is the Heaviside step function. For a projective measurement outcome $|\phi\>\< \phi|$ and a quantum state $|\psi\>$, the Born rule is satisfied since
\begin{eqnarray}
\int \d \lhat \d x \,\,\mu_{\psi} (\lhat ,x ) \xi_{\phi} (\lhat,x) = |\< \psi | \phi \> |^2.
\end{eqnarray}

Note---as Bell did---that there are many possible choices for the response functions that would work equally well. The only requirement is that the support over the subset $\{ (\hat\lambda, x): 0\le x \le 1 \}$ has measure equal to the probability occurring in the Born rule; how this support is distributed is entirely arbitrary.

\emph{A $\psi$-epistemic modification of the Bell model for a qubit.--}The Bell model is $\psi$-ontic, since no two epistemic states, corresponding to distinct quantum states, overlap. A different ontological model for qubit systems was described by Kochen and Specker \cite{Koc67}, which is $\psi$-epistemic. To date, however, no one has extended Kochen and Specker's model to systems of higher dimension than two \cite{Rud06}. This section shows how the Bell model can be modified in order to obtain a $\psi$-epistemic model for qubits. Later, this model is extended to obtain a $\psi$-epistemic model for quantum systems of arbitrary finite dimension.

\begin{figure}
\begin{center}
\setlength{\unitlength}{0.45\columnwidth}
\subfloat[]{
\begin{picture}(1,1)
\put(0.0,0.0){\includegraphics[width=0.45\columnwidth]{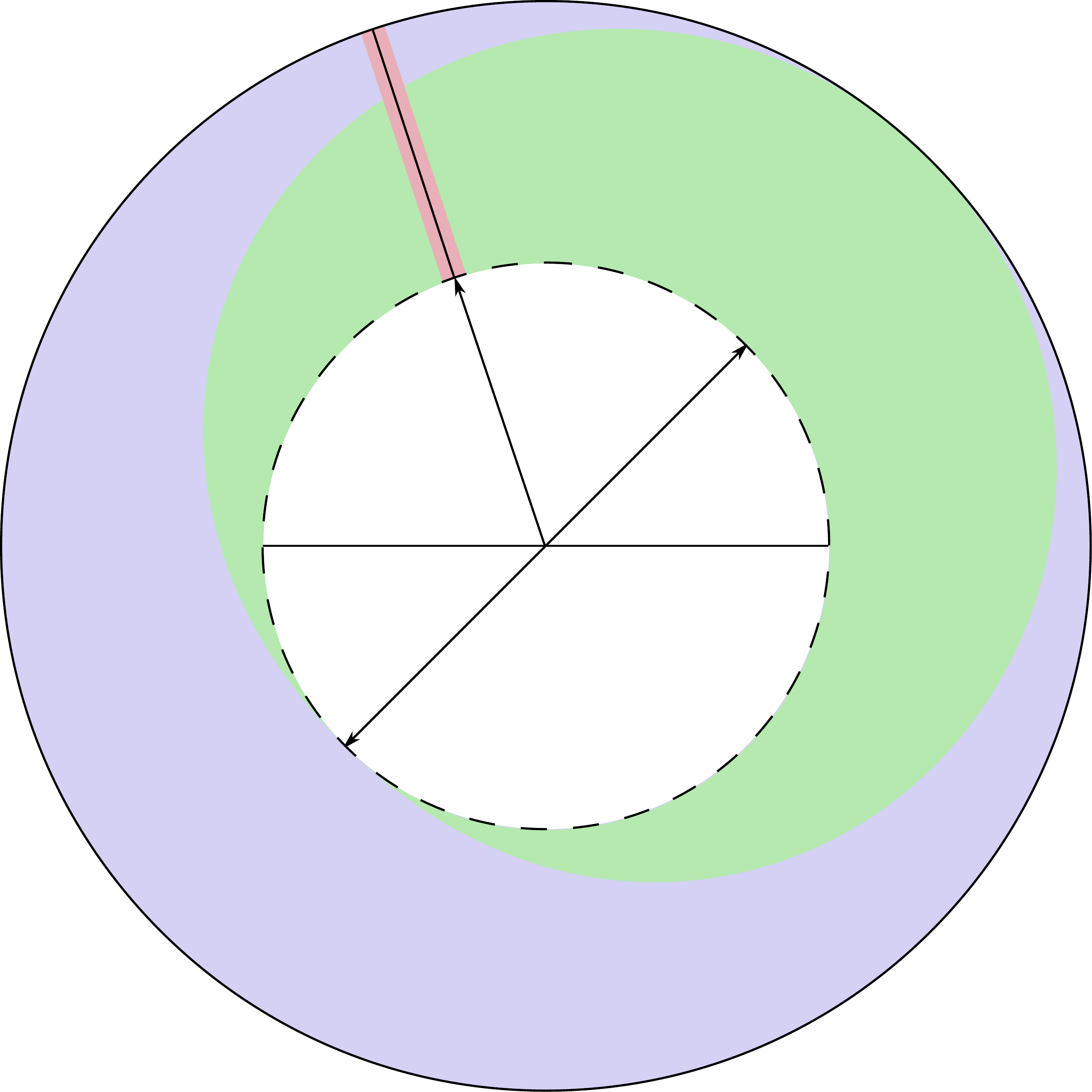}}
\put(0.64,0.58){\(\hat{\phi}_0\)}
\put(0.37,0.64){\(\hat{\psi}\)}
\put(0.45,0.8){\(\supp(\xi_{\phi_0})\)}
\put(0.25,0.1){\(\supp(\xi_{\phi_1})\)}
\put(0.36,0.3){\(\hat{\phi}_1\)}
\end{picture}
}
\subfloat[]{
\begin{picture}(1,1)
\put(0.0,0.0){\includegraphics[width=0.45\columnwidth]{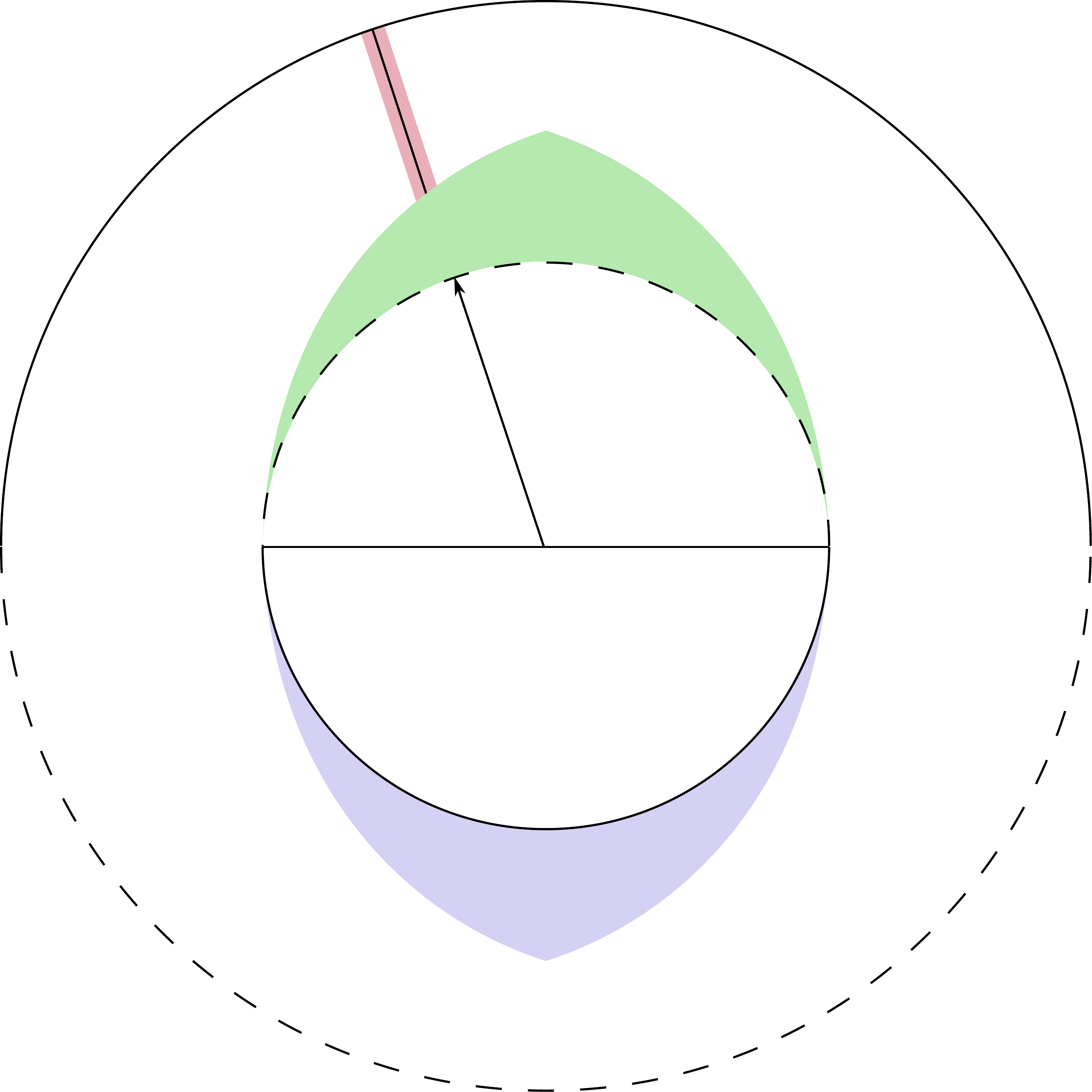}}
\put(0.37,0.64){$\psihat$}
\put(0.46,0.785){$\E_0$}
\put(0.46,0.173){$\E_1$}
\put(0.95,0.813){$\R_0$}
\put(0.95,0.12){$\R_1$}
\end{picture}
}
\caption{(a) Unmodified Bell model ontic state space along with supports of a response function for outcome $\hat{\phi_0}$ (green), $\hat{\phi_1}$ (blue) and epistemic state $\mu_{\psi}$ (red line). The inner circle is the surface of the Bloch sphere, the dashed line indicates it also corresponds to $x=0$. The outer solid circle corresponds to $x=1$. (b) Modified Bell model, showing the subsets $\E_0$ and $\E_1$. Any probability weight that $\mu_{\psi}$ gives to ontic states within $\E_0$ can be redistributed over $\E_0$.\label{annuli}}
\end{center}
\end{figure}
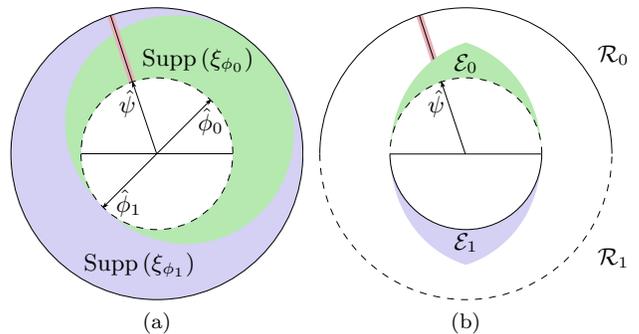

A simple visualization of the ontic state space $\Lambda$ as an annulus with $\lhat$ specifying the direction, and $x$ the radial distance is depicted in Fig.~\ref{annuli}(a). Let $\hat{z}$ correspond to the north pole of the Bloch sphere, and let $\lhat\cdot\hat{z} = \cos(\theta_\lambda)$, where \(\theta_\lambda\) is the polar angle of the Bloch vector \(\hat{\lambda}\). Label the upper ($\theta_\lambda < \pi/2$) hemisphere $\R_0$ and the lower ($\theta_\lambda > \pi/2$) hemisphere $\R_1$. Given a projective measurement $\Phi =\{ |\phi_0\>\< \phi_0|, |\phi_1\>\< \phi_1| \}$, assume that the outcomes are labelled such that $|\<\phi_0|z\>|^2 \geq |\<\phi_1|z\>|^2$.

With response functions defined as in Eq.~(\ref{BellIndicator}), the ontic states in the set
\begin{equation}
\E_0  = \left\{ (\lhat, x):  \lhat \in \R_0 \mbox{ and } 0\le x < (1 - \sin \theta_\lambda)/2 \right\}
\end{equation}
all result in the $\hat{\phi}_0$ outcome for any measurement $\Phi$.
Similarly, those in the set
\begin{equation}
\E_1  = \left\lbrace (\lhat, x):  \lhat \in \R_1 \mbox{ and } (1 + \sin \theta_\lambda)/2 < x \leq 1 \right\rbrace
\end{equation}
all result in the $\hat{\phi}_1$ outcome.  
These sets are illustrated in Fig.~\ref{annuli}(b).

In order to construct a $\psi$-epistemic model, note that if an epistemic state $\mu_{\psi}$ assigns non-zero probability to the subset $\E_0$, then this much probability weight can be redistributed over $\E_0$ without changing the Born rule statistics. This is because ontic states in $\E_0$ behave identically to one another, as far as predictions for quantum measurement outcomes go. Similarly $\E_1$. Hence define a modified Bell model such that for $\hat{\psi} \in \R_0$,
\begin{multline}
\mu_{\psi} (\lhat, x) = \delta(\lhat - \psihat) \Theta\left( x - \frac{1}{2} (1 - \sin \theta_\psi) \right) \\+ \frac{1}{2} (1 - \sin \theta_\psi) \mu_{\E_0} (\lhat , x)\nonumber
\end{multline}
where $\theta_\psi$ is the polar angle of $\psihat$, and $\mu_{\E_0}$ is essentially arbitrary but can be taken to be the uniform distribution over $\E_0$. A similar expression defines $\mu_{\psi}$ for states with $\psihat \in \R_1$.

The modified Bell model still reproduces the Born rule, but now the model is $\psi$-epistemic. Any two quantum states in the same hemisphere are described by distributions that overlap, either in $\E_0$ or in $\E_1$. The preparation of an ontic state from either of these regions does not reveal a unique quantum state, but only reveals in which hemisphere the quantum state resides.

\emph{$\psi$-epistemic modification in higher dimensions.--}Here, we modify the Bell model to produce a $\psi$-epistemic model for arbitrary finite dimension. The ontic state space for the $d$-dimensional Bell model is $\Lambda = CP^{d-1} \times [0,1]$. The distribution corresponding to a quantum state $|\psi\>$ is given by $\mu_{\psi} (\ket{\lambda}, x) = \delta (\ket{\lambda} -\ket{\psi})$. Response functions for a measurement $\Phi$ are defined such that $\xi_{\phi_k}$ has support of length $|\<\phi_k | \lambda\>|^2$ on the line segment $ \{ (\ket{\lambda}, x): 0 \le x \le 1\}$. Up to this constraint the response functions are arbitrary. This model is $\psi$-ontic since the delta functions do not overlap for distinct $\ket{\psi}$.

In order to construct a $\psi$-epistemic model, fix an arbitrary preferred state $|0\>$. For each measurement $\Phi$, assume that the outcomes are ordered such that $|\<\phi_0|0\>|^2 \geq |\<\phi_1|0\>|^2 \geq \cdots \geq |\<\phi_{d-1}|0\>|^2$. Fix the response functions so that
\begin{equation}\label{higherdresponse}
\xi_{\phi_k}(|\lambda\>,x) = 1 \quad \mathrm{if}\quad \sum_{i=0}^{k-1} |\<\lambda|\phi_i\>|^2 \leq x < \sum_{i=0}^{k} |\<\lambda | \phi_i\rangle|^2 
\end{equation}
and 
\begin{equation}
\xi_{\phi_k}(|\lambda\>,x) = 0 \quad \mathrm{otherwise}.
\end{equation}
In Equation~(\ref{higherdresponse}), $\sum_{i=1}^{k-1}|\<\lambda|\phi_i\>|^2$ is taken to be $0$ when $k=0$. For completeness, let us specify that for $x=1$, $\xi_{\phi_k}(|\lambda\>,x) = 1$ iff $k = d-1$.

Now the aim is to define a subset $\E_0$ of $\Lambda$, such that ontic states in $\E_0$ predict the same outcomes for all measurements. To this end, note that $|\<\phi_0|0\>|^2 \geq 1/d$. Given this, it is easy to show that if $|\<\lambda|0\>|^2 > \frac{d-1}{d}$, then $|\<\phi_0|\lambda\>|^2 > 0$. For arbitrary $|\chi\>$, let 
\begin{equation}
z(|\chi\>) = \inf_{|\phi\>: |\<\phi|0\>|^2 \geq 1/d } |\<\phi|\chi\>|^2,
\end{equation}
where an explicit expression is easily found but not needed.
Define
\begin{equation}
\E_0  = \left\{ (|\lambda\>, x):  |\<\lambda|0\>|^2 > \frac{d-1}{d} \mbox{ and } 0\le x < z(|\lambda\>) \right\}.
\end{equation}
Any ontic state $\lambda \in \E_0$ has the property that whatever measurement is performed, the outcome is $|\phi_0\>\<\phi_0|$. The epistemic states can therefore be modified as above to produce a $\psi$-epistemic model. Informally, the idea is the same: any probability that $\mu_{\psi}$ assigns to ontic states within the set $\E_0$ can be redistributed over the whole of $\E_0$ without changing Born rule statistics. More specifically, when $|\<\psi|0\>|^2 \leq (d-1)/d$, let
\begin{equation}
\mu_{\psi}(|\lambda\>,x) = \delta(\ket{\lambda} -\ket{\psi}),
\end{equation}
and when $|\<\psi|0\>|^2 > (d-1)/d$, let
\begin{equation}
\mu_{\psi}(|\lambda\>,x) = \delta(\ket{\lambda} - \ket{\psi}) \Theta( x - z(|\psi\>)) + z(|\psi\>) \mu_{\E_0} (|\lambda\> , x),
\end{equation}
where, as above, $\mu_{\E_0}$ is arbitrary but could be taken to be the uniform distribution over $\E_0$.

This model is clearly very contrived. A degree of symmetry could be restored by postulating a preferred basis $|0\rangle,\ldots,|d-1\rangle$, instead of a preferred state. If $|\<\lambda|j\>|^2 > (d-1)/d$ for some $j$, then relabel measurement outcomes such that $|\<\phi_0|j\>|^2 \geq |\<\phi_1|j\>|^2 \geq \cdots \geq |\<\phi_{d-1}|j\>|^2$. In this case $|\<\phi_0|\lambda\>|^2 > 0$, and sets $\E_j$ can be defined in analogy with $\E_0$ above. For any state $|\psi\rangle$ with $|\<\psi|j\>|^2 > (d-1)/d$, any probability assigned to ontic states within $\E_j$ can be redistributed over $\E_j$. This results in a model that is `more epistemic' than the one above. It is an open question whether more natural models can be found. 

\emph{Discussion.--}Our results are particularly pertinent given the recent no-go theorem of \cite{PBR}. The theorem shows that it is not possible to construct $\psi$-epistemic models of quantum theory, given an assumption that independent preparations produce uncorrelated ontic states. The present Letter shows that $\psi$-epistemic models are possible if that assumption is given up. None of these models is intuitive or motivated by physical principles or considerations. The primary motivation for exploring the possibility of $\psi$-epistemic models is to understand the formal limitations of reproducing quantum theory from a deeper theory. 

Such models may also play a useful role in other areas of quantum information. The remarkable protocols of Toner and Bacon \cite{Ton03} for recovering the bipartite measurement statistics of a singlet state and classical simulation of teleportation---using only one and two bits (respectively) of classical communication---make use of shared randomness in a way that lets the problem be recast in terms of correlated single qubit ontological models that are $\psi$-epistemic. A related result has been obtained by Montina \cite{Mon11}, who has also put bounds on the classical simulation cost for an arbitrary number of qubits.

There remain many open questions. In the models presented, it is not the case that $\mu_{\psi}$ has non-zero overlap with $\mu_{\phi}$ for any pair of non-orthogonal quantum states $|\psi\rangle$ and $|\phi\rangle$. Hence the models do not satisfy the stronger definition of $\psi$-epistemic suggested above. It would be interesting to establish whether such models exist. (Scott Aaronson has recently combined some of the ideas that were presented in a preprint version of this article with those of George Lowther to answer this question in the affirmative; see \cite{Aar12}.)

The authors gratefully acknowledge support from the Engineering and Physical Sciences Research Council (P.G.L., J.B., and T.R.), the Leverhulme Foundation (T.R.) and the Royal Comission for the Exhibition of 1851 (D.J.).

\end{document}